\documentclass[letterpaper, 10 pt, journal, twoside]{IEEEtran}
\usepackage{amsmath,amssymb,amsbsy,enumerate,multirow,subfigure,epsfig,hhline,multirow,srcltx,float}
\usepackage[usenames]{color}
\usepackage[ruled, lined, linesnumbered]{algorithm2e}
\newtheorem{theorem}{Theorem}

\newtheorem{assumption}{Assumption}
\newtheorem{remark}{Remark}

\allowdisplaybreaks

\DeclareMathOperator*{\argminC}{\arg\min}   

\begin{document}

\title{Red Light, Green Light Game of Multi-Robot Systems
	with Safety Barrier Certificates}

\author{Yun Ho Choi and Doik Kim,~\IEEEmembership{Member,~IEEE}
\thanks{
This work was supported by the Korea Institute of Science and Technology (KIST) Institutional Program under Grant 2E31581 and by Basic Science Research Program through the National Research Foundation of Korea(NRF) funded by the Ministry of Education(NRF-2021R1A6A3A01086607).
(\textit{Corresponding author: Doik Kim.}}
\thanks{The authors are with the Center for Intelligent \& Interactive Robotics Research,
	Korea Institute of Science and Technology, Seoul, 02792, South Korea (e-mail: yhchoi@kist.re.kr; doikkim@kist.re.kr).}
}

\markboth{IEEE ROBOTICS AND AUTOMATION LETTERS. PREPRINT VERSION. xxx, xxx} 
{CHOI AND KIM: Red Light, Green Light Game of Uncertain Multi-Robot Systems with Safety Barrier Certificates}

\maketitle

 \begin{abstract}
	In this paper, we propose the safety barrier certificates for uncertain multi-robot systems playing red light, green light game. 
	According to the rule of the game, the robots are allowed to move forward after a doll shouts `green light' and must stop when it shouts `red light'. Following this rule, a two-mode nominal controller is designed where one mode is for moving forward and the other one is for slowing down and being motionless. Then, multiple exponential control barrier functions(ECBFs) are developed to handle safety constraints for limited playground, collision avoidance, and saturation of the velocity. 
	While designing the nominal controller and ECBFs, an estimated braking time and robust inequality constraints are derived to deal with the system uncertainty.
	Consequently, a controller guaranteeing safety barrier certificates of each robot has been formulated by a quadratic programming with the nominal controller and the robust inequality constraints. 
	Finally, red light, green light game is simulated to validate the proposed safety-critical control system.
\end{abstract}


\section{Introduction}
\IEEEPARstart{E}{nsuring} safety is critical for many applications such as autonomous vehicles, chemical plants, robotic systems, and so on. Regarding safety verification as a forward invariance of a safe subset of the state space, 
control barrier functions (CBFs) are proposed to design control strategies ensuring that the system states stay in the safe set \cite{Ames2017}. The main concept of CBFs is to map the safety constraints over system states onto a constraint on the control input.
Lots of researchers have addressed safety-critical control using CBFs of practical applications like Segway, bipedal walking robots, and quadrotors \cite{Gurriet2018,Hsu2015,Wu2016} and have designed new CBFs for handling nonlinear systems with high-relative degrees, hybrid nonlinear systems, and uncertain nonlinear systems \cite{Nguyen2016,Nguyen202x,Lopez2021}.

On the other hand, cooperative control of multi-robot systems has been extensively studied in order to realize complicated tasks, such as environmental surveillance, disaster rescue, and minefield mapping \cite{Khamis2015,Smith2012}.
One of the main challenges in operating multi-robot systems is collision avoidance among robots. 
To resolve this issue, potential function approach is exploited in \cite{Mastellone2008,Li2013,Peng2020} and 
reciprocal velocity obstacle technique is adopted in \cite{Berg2009,Snape2011}. 
In addition, model predictive control is known as an efficient way to guarantee collision avoidance as reported in \cite{Richards2004,Dai2017}. For further references, see a survey paper by Hoy \cite{Hoy2014}. 
Owing to the advantage of CBFs in terms of scalability and realtimeness \cite{Li202x} compared to the aforementioned approaches, CBFs-based control designs for constrained multi-robot systems have been presented \cite{Wang2017,Emam2019,Luo2020,Chen2021}. 
In \cite{Wang2017}, safety barrier certificates for collision avoidance of double-integrator multi-robot systems have been presented where safety barrier constraints among all pairs of robots are defined involving the relative braking distance computed from the limited acceleration. 
This strategy has been extended for disturbed multi-robot systems by formulating robust barrier functions 
and its effectiveness was successfully verified by a long-term experiment in the Robotarium \cite{Emam2019}.
Lue et al.,\cite{Luo2020} provided probabilistic safety barrier certificates in the presence of process and measurement noise in collision/obstacle avoidance. Recently, CBF with one or multiple backup controllers was proposed in \cite{Chen2021} 
where the backup controllers are introduced to get a feasible motion of multiple robots satisfying the safety constraint.

Meanwhile, Squid Game \cite{Young2021}, Netflix's drama series, has earned incredible international success since it was released.
In this show, 456 players risk their lives to play a series of six children's games for the chance to win the prize money. 
Among the six games, the first game is red light, green light. 
The goal of the players in this game is to cross the finish line before the timer runs out and the players are allowed to move forward after a doll shouts `green light' and they have to stop when it shouts `red light'. If movement is detected afterward, they are eliminated. Since all of the players' characters are not the same, they act differently e.g., some of them act conservatively because of the fear of being eliminated and others act aggressively to cross the line in time.
On the other hand, control of connected vehicles with vehicle-to-vehicle and vehicle-to-infrastructure communications become a significant topic of intelligent transportation systems. One of the main challenges is to consider connected vehicles in urban road conditions with congested traffic and traffic lights \cite{HC2016, HC2017, He2019}. 
In these studies, a desirable distance between a vehicle and its preceding vehicle is defined and a target velocity is chosen wisely with least vehicle idling at red lights. Then, optimal longitudinal control strategies were developed to keep the desirable distance similar to the collision avoidance of multi-robot systems and to follow the target velocity. 
Notice that red light, green light game of multi-robot systems is an extended version of control problem of connected urban vehicles with traffic signal because it is played in 2D playground and robots can pass slow-moving robots with guaranteed safety.

Motivated by this observation, this paper addresses red light, green light game of multi-robot systems. 
To reflect realistic and heterogeneous players, viscous friction is assumed to be present and 
the saturation values of the velocity and acceleration of each robot are set to be different from each robot.   
Following the game rule, a two-mode nominal controller is developed for each robot where one is for moving forward and the other is for slowing down and being motionless. Multiple safety constraints for limited playground, collision avoidance, and saturation of the velocity in the game are encoded by exponential CBFs. 
Since there exist viscous friction, an estimated braking time is computed instead of the real braking time in developing the nominal controller and robust inequality constraints on the control input are derived.
Consequently, the control input guaranteeing safety barrier certificates of each robot has been formulated by a quadratic programming with the nominal controller and those robust inequality constraints. 
To demonstrate the effectiveness of the proposed safety-critical control system during the play, simulation results are presented.

The rest of this paper is organized as follows. Section II presents the background of exponential CBFs and formulates the control problem for realizing red light, green light. Section III gives some main results including the nominal controller and the safe  controller design in the presence of system uncertainty and some constraints.
In Section IV, simulation results are shown to validate the effectiveness of the proposed safety-critical control system for the game. Section V offers some conclusions.

\section{PROBLEM FORMULATION AND BACKGROUND}
\subsection{Problem statement}
The players of red light, green light game are modeled as uncertain double-integrator robots as follows:
\begin{align}
	\begin{array}{l}
		\dot{p}_{i} = v_{i}, \\
		\dot{v}_{i} = u_{i} - \kappa_{i}v_{i}
	\end{array}
	\label{eq:robot}
\end{align}
where $i \in \mathcal{N}$; $\mathcal{N} =\{i ~| i=1,\dots,N \}$, $p_{i} = [p_{i,x}, p_{i,y}]^{\top}$, $v_{i} = [v_{i,x}, v_{i,y}]$, and $u_{i} = [u_{i,x}, u_{i,y}]$ denote the position, velocity, and acceleration of the $i$th robot and the term $-\kappa_{i}v_{i}$ indicates the viscous friction of the robot. 
The velocities and accelerations are saturated by $-V_{i} \leq v_{i,x}, v_{i,y} \leq V_{i}$ and $-U_{i} \leq u_{i,x},u_{i,y} \leq U_{i}$ where $V_{i}$ and $U_{i}$ are the saturation values. Note that these values are different from each robot to reflect the heterogeneity of each player. 
In this paper, we use the following four assumptions.

\begin{assumption}
	\label{as:friction}	
	The friction constants $\kappa_{i}$ are unknown but satisfy $\kappa_{low} \leq \kappa_{i} \leq \kappa_{up}$ with known positive constants $\kappa_{low}$ and $\kappa_{up}$.
\end{assumption}

\begin{assumption}
	\label{as:radius}
	Each robot is modeled as a circular disk of radius $r_{0}$.
\end{assumption}

\begin{assumption}
	\label{as:red-green-time}
	Let $t_{g_{k}}$ and $t_{r_{k}}$ be the moments when the doll shouts green light and red light, respectively, with $k=1,2,\dots, \bar{k}$. Here, $\bar{k}$ is a positive constant, $t_{g_{1}} = 0$, and $t_{r_{\bar{k}}} = +\infty$.
	During the play, the moments $t_{g_{k}}$ and $t_{r_{k}}$ are known for every robot. 
\end{assumption}

\begin{assumption}
	\label{as:playground}
	The playground $\mathcal{P}$ of the robots is a fixed, closed, and bounded rectangular region defined as $\mathcal{P} = \{p_{i}\in \mathbb{R}^{2}: 0 \leq p_{i,x} \leq L_{x}, 0 \leq p_{i,y} \leq L_{y}\}$. 
	For all robots, $L_{x}>0$ and $L_{y}>0$ are known.
\end{assumption}

The objective of this paper is to design a safety-critical controller for each robot playing red light, green light game under Assumptions \ref{as:friction}--\ref{as:playground} 
so that each robot crosses the finish line while avoiding collision, and remaining within the playground.

\subsection{Exponential control barrier functions(ECBFs)}
To handle some high relative degree constraints of the robots playing red light, green light, 
we use ECBFs introduced in \cite{Nguyen2016}. 
Along Definition 1 in \cite{Nguyen2016}, a function $H: \mathbb{R}^{n} \rightarrow \mathbb{R}$ is an ECBF of relative degree two for a dynamical system $\dot{x} = f + gu$ if there exists $\gamma>0$ such that
\begin{align}
	\nonumber 
	&\sup_{u}[L_{f}^{2}H(x) + L_{g}L_{f}H(x)u 
	\\ 
	& \qquad  + \gamma^{2} H(x) +  2\gamma L_{f}H(x) ] \geq 0
	\label{eq:cond-ECBF}
\end{align}
for $\forall x \in \{ x \in \mathbb{R}^{n} | H(x_{0}) \geq 0, L_{f}H(x_{0})+\gamma H(x_{0}) \geq 0\}$ that yields
$H(x(t)) \geq 0 $, $\forall t \geq 0$. For more general information of ECBFs, please see \cite{Nguyen2016}. 

A controller that satisfies \eqref{eq:cond-ECBF} renders the system $\dot{x} = f + gu$ safe with respect to a set $\mathcal{S} \triangleq \{x \in \mathbb{R}^{n} | H(x) \geq 0 \}$. Thus, \eqref{eq:cond-ECBF} is used for deriving a local control law $u_{i}$ of the uncertain robot \eqref{eq:robot} with multiple safety constraints in the next section.

\section{MAIN RESULTS}
\subsection{Nominal controller design}
\begin{figure}
	\centering
	\includegraphics[width=8cm,height=2.5cm]{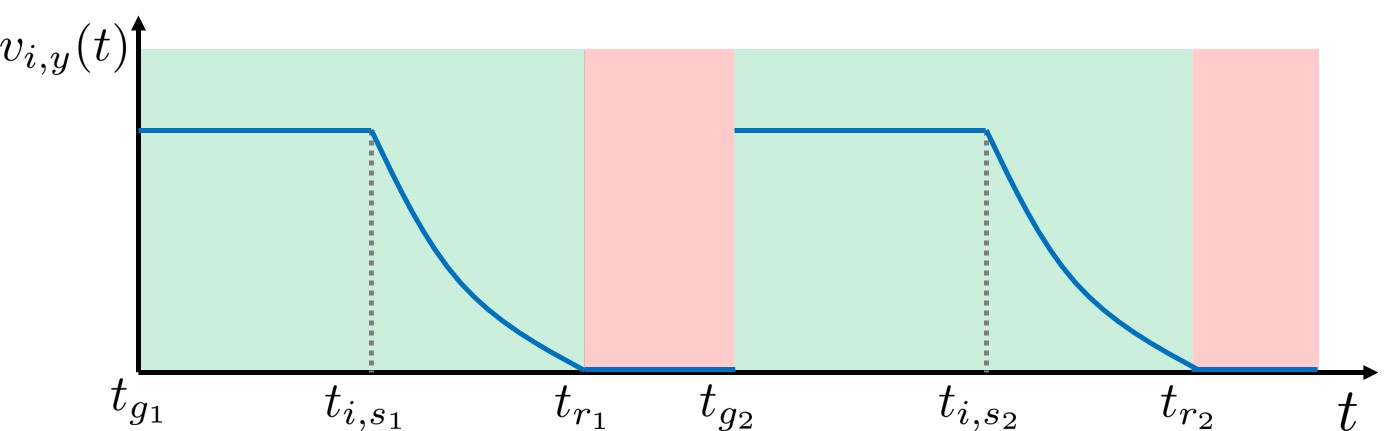}
	\caption{Example history of the robot's velocity on y-axis.}
	\label{fig:Exam_velocity}
\end{figure}

In this section, we present a nominal controller for each robot. 
From the rule of the game, the robots can only move for $t \in [t_{g_{k}}, t_{r_{k}})$ 
and they are eliminated if their movement is detected for $t \in [t_{r_{k}}, t_{g_{k+1}})$ where $k=1,2,\dots, \bar{k}$. 
Thus, the robots should move forward to cross the finish line as fast as they can, 
but they should slow down for a certain period of time to be motionless at $t = t_{r_{k}}$ as illustrated in Fig. \ref{fig:Exam_velocity}.
Let $t_{i,s_{k}}$ be the start time to slow down of the $i$th robot such that $t_{g_{k}} \leq t_{i,s_{k}} < t_{r_{k}}$. In this paper, $t_{i,s_{k}}$ is defined as
\begin{align}
	t_{i,s_{k}} = \{ t \geq t_{g_{k}} | ~ t_{r_{k}}- t = \eta_{i}\hat{t}_{i}^{*}\}
	\label{eq:stop-time}
\end{align}
where $\eta_{i} \geq 1$ is a design parameter describing the sensitivity to the rule and $\hat{t}_{i}^{*}$ is an estimated braking time. To compute $\hat{t}_{i}^{*}$, let us assume that the current velocity on y axis is positive and the maximum deceleration is activated (i.e., $v_{i,y}(t_{0})>0$ and $u_{i,y}(t) = -U_{i}$). 
Then, solving $u_{i,y}(t) = -U_{i}  - \kappa_{i}v_{i,y}(t)$ from \eqref{eq:robot}, 
$v_{i,y}(t)$ for $t \geq t_{0}$ is given by 
\begin{align}
	\nonumber 
	v_{i,y}(t) & = -\bigg(1 - \mathrm{exp}(-\kappa_{i}(t-t_{0}))\bigg)\frac{U_{i}}{\kappa_{i}} 
	\\
	& \quad + \mathrm{exp}(-\kappa_{i}(t-t_{0}))v_{i,y}(t_{0}).
\end{align}

Let $t_{i,y}^{*}$ be the actual braking time on y-axis satisfying $v_{i,y}(t_{0}+t_{i,y}^{*}) = 0$. 
Then, by taking log on both sides, $t_{i,y}^{*}$ is 
computed by
\begin{align}
	t_{i,y}^{*} = \frac{1}{\kappa_{i}} \log\bigg(\frac{U_{i}/\kappa_{i} + v_{i,y}(t_{0})}{U_{i}/\kappa_{i}}\bigg).
\end{align}
However, $t_{i,y}^{*}$ is not implementable because of the unknown parameter $\kappa_{i}$. 
Thus, using $\kappa_{low} \leq \kappa_{i}$ from Assumption \ref{as:friction}, 
an estimated braking time on y-axis $\hat{t}_{i,y}^{*} \geq t_{i,y}^{*}$ is defined as
\begin{align}
	\hat{t}_{i,y}^{*} = \frac{1}{\kappa_{low}} \log\bigg(\frac{U_{i}/\kappa_{low} + v_{i,y}(t_{0})}{U_{i}/\kappa_{low}}\bigg).
\end{align}
Finally, $\hat{t}_{i}^{*}$ is defined as 
\begin{align}
	\hat{t}_{i}^{*} = \max\{ \hat{t}_{i,x}^{*}, \hat{t}_{i,y}^{*}\}.
	\label{eq:est-braking}
\end{align}

\begin{figure}
	\centering
	\subfigure[]{\epsfig{figure=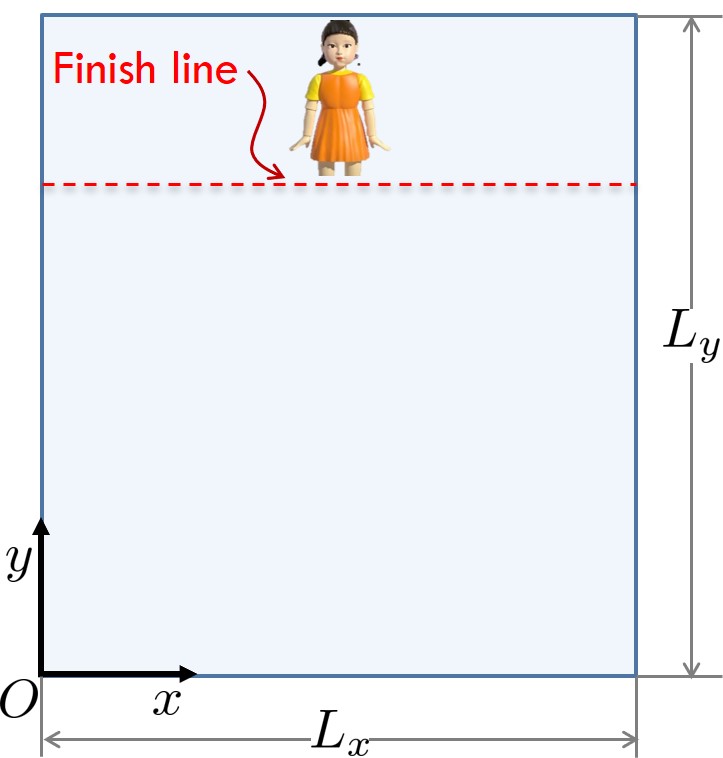, width=3.5cm,height=3.5cm}}
	\subfigure[]{\epsfig{figure=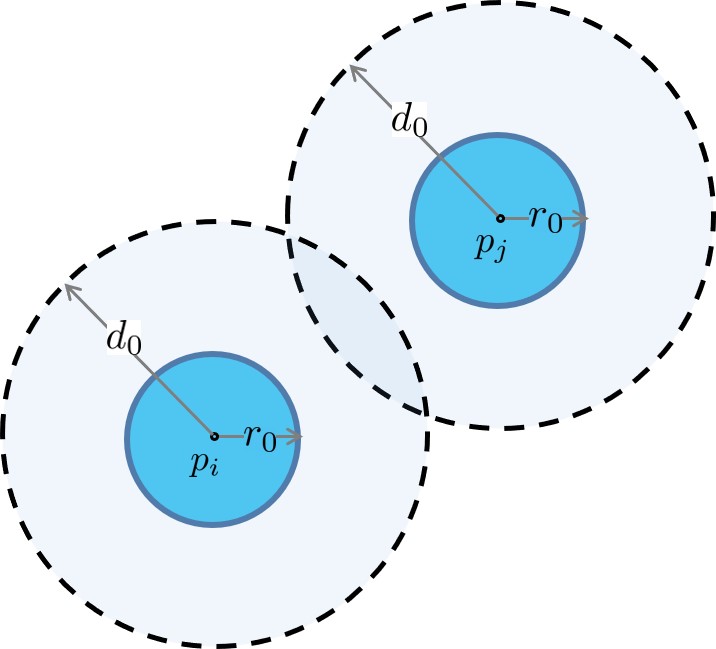, width=3.5cm,height=3.2cm}}
	\subfigure[]{\epsfig{figure=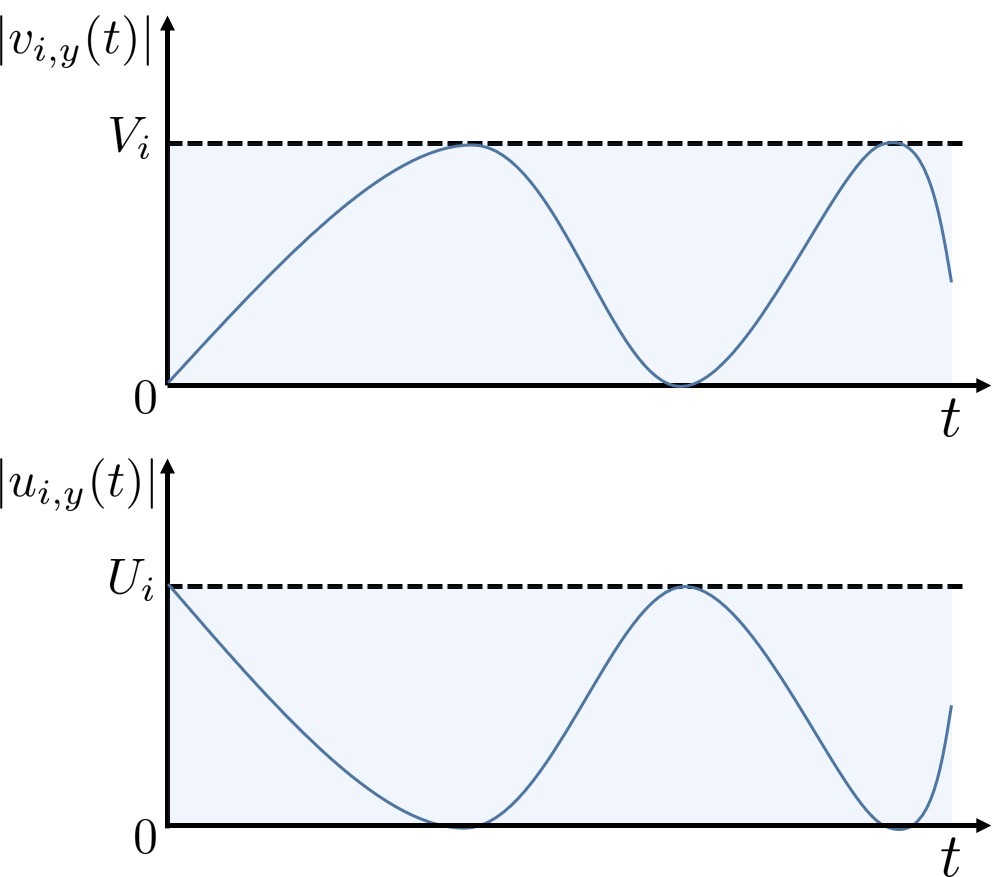, width=5cm,height=4cm}}
	\caption{Three safety constraints of the considered multi-robot systems
		(a) playground constraint
		(b) collision avoidance
		(c) saturation of the velocity and acceleration.}
	\label{fig:const}
\end{figure}

Computing \eqref{eq:est-braking} at every time and defining $t_{i,s_{k}}$ by  \eqref{eq:stop-time}, a two-mode nominal controller $u_{i,n} = [u_{i,n,x}, u_{i,n,y}]^{\top}$ is designed as follows:
\begin{align}
	\nonumber 
	u_{i,n,x}(t) &= \left\{
	\begin{array}{l}
		-k_{i}v_{i,x}(t), \qquad \quad ~~t \in [t_{g_{k}}, t_{i,s_{k}})\\
		-U_{i}\mathrm{sign}(v_{i,x}(t)), \quad t \in [t_{i,s_{k}}, t_{g_{k+1}})
	\end{array}
	\right.
	\\	
	u_{i,n,y}(t) &= \left\{
	\begin{array}{l}
		U_{i}, \qquad \qquad  \qquad \quad t \in [t_{g_{k}}, t_{i,s_{k}})\\
		-U_{i}\mathrm{sign}(v_{i,y}(t)), \quad t \in [t_{i,s_{k}}, t_{g_{k+1}})
	\end{array}
	\right.
	\label{eq:nomin-controller} 
\end{align}
where $k=1,2,\dots,\bar{k}$ and $k_{i}>0$ denotes the control gain. Note that $u_{i,n,y}(t) = U_{i}$ for $t \in [t_{g_{k}}, t_{i,s_{k}})$ because the goal of the robot is to cross the finish line (i.e., a horizontal line in the playground in Fig. \ref{fig:const}(a)).

\begin{remark}
	For the sensitivity parameter $\eta_{i}$ in \eqref{eq:stop-time}, it should be pointed out that
	\\
	\indent (i) we consider different parameters $\eta_{i}$ for different robots to reflect the heterogeneity of the game players;
	\\
	\indent (ii) it is not guaranteed that the robots do not violate the rule (i.e., being motionless for $t \in [t_{r_{k}}, t_{g_{k+1}})$) because the controller $u_{i}$ could be different from the ideal nominal controller $u_{i,n}$ owing to some safety constraints;
	\\	
	\indent (iii) if some $\eta_{i}$ are close to $1$, one can see that those robots are willing to take a risk to move as much as they can. Thus, such robots can represent `drastic' players.
	On the contrary, some robots with large $\eta_{i}$ will take conservative action by decelerating earlier which can stand for `careful' players.
\end{remark}

\begin{remark}
	Different from \cite{Wang2017}, the nominal controller of the robot is switched whenever $t = t_{i,s_{k}}$ because the robots have to decrease their velocity to not break the rule. 
	Since there exists a friction uncertainty in \eqref{eq:robot}, we have designed a nominal controller \eqref{eq:nomin-controller} using an estimated braking time $\hat{t}_{i}^{*}$ instead of the actual braking time $t_{i}^{*}$. 
\end{remark}

\begin{remark}
	Normally, the sign function in the controller often causes chattering phenomena of the solution because of digital implementation. Thus, $-U_{i}\mathrm{sign}(v_{i,m}(t))$ with $m=x,y$ can be replaced with $-U_{i}\mathrm{tanh}(\varepsilon v_{i,m}(t))$ to attenuate the chattering where $\varepsilon$ is a large constant. 
\end{remark}

\subsection{Safety-critical controller design}
In this section, a safety-critical controller of each robot will be designed. 
In Fig. \ref{fig:const}, three safety constraints for the robots are displayed during the play of red light, green light game.
To deal with these constraints, ECBFs are exploited. 

\emph{First constraint}: To maintain the movement of the robots of radius $r_{0}$ within the playground, 
two constraints for x- and y-axes can be expressed by
\begin{align}
	\begin{array}{l}
		1.1r_{0} \leq p_{i,x} \leq  L_{x} - 1.1r_{0}, \\
		1.1r_{0} \leq p_{i,y} \leq  L_{y} - 1.1r_{0}. \\		
	\end{array}
	\label{eq:const1}
\end{align}
Notice that we use $1.1r_{0}$ instead of $r_{0}$ to prevent the robots from contacting the boundaries of the playground. Since \eqref{eq:const1} are high relative degree safety constraints, exponential CBFs \cite{Nguyen2016} are designed as
\begin{align}
	\begin{array}{l}
		h_{i,1,x} = ( \bar{L}_{x} - 1.1r_{0} )^{2} - (p_{i,x} - \bar{L}_{x} )^{2}, \\
		h_{i,1,y} = (\bar{L}_{y} - 1.1r_{0} )^{2} - (p_{i,y} - \bar{L}_{y} )^{2}.
	\end{array}
	\label{eq:CBF1}
\end{align}
where $\bar{L}_{x} = L_{x}/2$ and $\bar{L}_{y} = L_{y}/2$. The first and second time derivatives of $h_{i,1,x}$ using \eqref{eq:robot} are given by 
\begin{align}
	\dot{h}_{i,1,x} & = - 2(p_{i,x} - \bar{L}_{x})v_{i,x}, \\
	\ddot{h}_{i,1,x} & = -2v_{i,x}^{2} - 2(p_{i,x} - \bar{L}_{x})(u_{i,x} - \kappa_{i}v_{i,x}).
	\label{eq:ddot-h1}
\end{align}

Considering the uncertain robot model \eqref{eq:robot} and \eqref{eq:ddot-h1}, \eqref{eq:cond-ECBF} becomes 
\begin{align}
	\nonumber
	& -2v_{i,x}^{2} - 2(p_{i,x} - \bar{L}_{x})(u_{i,x} - \kappa_{i}v_{i,x})
	\\ 
	& \qquad  + \gamma_{i,1}^{2}h_{i,1,x} + 2\gamma_{i,1}\dot{h}_{i,1,x} \geq 0
	\label{eq:const1-ECBF-cond}
\end{align}
where $\gamma_{i,1}>0$ is a design parameter. By rearranging it, we have 
\begin{align}
	\nonumber 
	2(p_{i,x} - \bar{L}_{x})u_{i,x} & \leq 2(p_{i,x} - \bar{L}_{x})\kappa_{i}v_{i,x}
	-2v_{i,x}^{2} 
	\\
	& \quad + \gamma_{i,1}^{2}h_{i,1,x} + 2\gamma_{i,1}\dot{h}_{i,1,x}.
	\label{eq:const1-ECBF-cond2}
\end{align}
Since $\kappa_{i}$ is unknown, the above inequality is not implementable. 
To handle this issue, we use Assumption \ref{as:friction} to derive the inequality
\begin{align}
	2(p_{i,x} - \bar{L}_{x})\kappa_{i} v_{i,x}
	\geq - 2 \kappa_{up} |(p_{i,x} - \bar{L}_{x}) v_{i,x}|.
	\label{eq:const1-friction}
\end{align}
Then, applying \eqref{eq:const1-friction} into \eqref{eq:const1-ECBF-cond2}, an available robust constraint is obtained as
\begin{align}
	A_{i,1,x}u_{i,x} & \leq b_{i,1,x}
\end{align}
where
\begin{align}
	\nonumber
	A_{i,1,x} &= 2(p_{i,x} - \bar{L}_{x}), 
	\\ \nonumber
	b_{i,1,x} &= - 2 \kappa_{up}  |(p_{i,x} - \bar{L}_{x})v_{i,x}| 
	-2v_{i,x}^{2} 
	\\ \nonumber 
	& \quad + \gamma_{i,1}^{2}h_{i,1,x} + 2\gamma_{i,1}\dot{h}_{i,1,x}.
\end{align}
Similarly, an inequality constraint on $u_{i,y}$ is also obtained. 
Then, the first constraint on $u_{i}$ from the playground is given by
\begin{align}
	A_{i,1}u_{i} \leq b_{i,1}
	\label{eq:first-const}
\end{align}
where 
\begin{align}
	\nonumber
	A_{i,1} = 
	\left[
	\begin{array}{cc}
		A_{i,1,x} & 0 \\
		0 & A_{i,1,y}
	\end{array}
	\right], 
	\qquad 
	b_{i,1}  = \left[
	\begin{array}{c}
		b_{i,1,x} \\
		b_{i,1,y}
	\end{array}
	\right]. 
\end{align}

\emph{Second constraint}: To avoid collision among robots, the relative distance $d_{ij} = \| p_{i} - p_{j} \|$ between a pair of robots should satisfy 
\begin{align}
	d_{ij} \geq d_{0}, \qquad i,j \in \mathcal{N} ~~ i\neq j
	\label{eq:const2}
\end{align}
where $d_{0}>r_{0}$ indicates the avoidance region. 
An ECBF candidate for \eqref{eq:const2} is defined as 
\begin{align}
	h_{ij,2} = d_{ij} - d_{0}. 
\end{align}
Then, $\dot{h}_{i,j,2}$ and $\ddot{h}_{i,j,2}$ are given by
\begin{align}
	\label{eq:h2-dot}
	\dot{h}_{ij,2} & = \frac{p_{ij}^{\top}v_{ij}}{d_{ij}}, 
	\\ \label{eq:h2-ddot}
	\ddot{h}_{ij,2} & = \frac{\|v_{ij} \|^{2} + p_{ij}^{\top}u_{ij} - p_{ij}^{\top}(\kappa_{i}v_{i} - \kappa_{j}v_{j})}{d_{ij}} - \frac{(p_{ij}^{\top}v_{ij})^{2}}{d_{ij}^{3}}
\end{align}
where $p_{ij} = p_{i} - p_{j}$ and $v_{ij} = v_{i} - v_{j}$.
From \eqref{eq:h2-dot} and \eqref{eq:h2-ddot}, \eqref{eq:cond-ECBF} is given by 
\begin{align}
	\nonumber 
	-\frac{p_{ij}^{\top}u_{ij} }{d_{ij}}
	& \leq \frac{\|v_{ij} \|^{2}  - p_{ij}^{\top}(\kappa_{i}v_{i} - \kappa_{j}v_{j})}{d_{ij}} - \frac{(p_{ij}^{\top}v_{ij})^{2}}{d_{ij}^{3}}
	\\
	& \quad 
	+ \gamma_{i,2}^{2}h_{ij,2} + 2\gamma_{i,2}\dot{h}_{ij,2}
	\label{eq:const2-ECBF-cond}
\end{align}
where $u_{ij} = u_{i} - u_{j}$. Multiplying both sides of \eqref{eq:const2-ECBF-cond} by $d_{ij}$ and using the following property
\begin{align}
	\nonumber 
	& - p_{ij}^{\top}(\kappa_{i}v_{i} - \kappa_{j}v_{j}) 
	\\ \nonumber 
	& \qquad  = - \kappa_{i} p_{ij}^{\top}v_{ij} +  (\kappa_{j}-\kappa_{i}) p_{ij}^{\top}v_{j}
	\\ 
	& \qquad \quad   \geq - \kappa_{up} |p_{ij}^{\top}v_{ij}| - (\kappa_{up}-\kappa_{low})|p_{ij}^{\top}v_{j}|,
	\label{eq:const2-friction}
\end{align}
a coupled pairwise constraint is derived as follows:
\begin{align}
	A_{ij,2}u_{ij} \leq b_{ij,2}
	\label{eq:second-const-coupled}
\end{align}
where
\begin{align}
	\nonumber 
	A_{ij,2} & = -p_{ij}^{\top}, 
	\\ \nonumber
	b_{ij,2} & =  - \kappa_{up} |p_{ij}^{\top}v_{ij}| - (\kappa_{up}-\kappa_{low})|p_{ij}^{\top}v_{j}|+ \|v_{ij} \|^{2} 
	\\ \nonumber 
	& \quad 
	- (p_{ij}^{\top}v_{ij})^{2}/d_{ij}^{2} + \gamma_{i,2}^{2}d_{ij}h_{ij,2} + 2\gamma_{i,2}d_{ij}\dot{h}_{ij,2} 
\end{align}

In order to improve the scalability, the coupling pariwise constraints \eqref{eq:second-const-coupled} are divided into the two distributed constraints as follows \cite{Wang2017}:
\begin{align}
	\nonumber 
	A_{ij,2}^{\top}u_{i} \leq \frac{U_{i}}{U_{i}+U_{j}}b_{ij,2},  \quad 
	-A_{ij,2}^{\top}u_{j} \leq \frac{U_{j}}{U_{i}+U_{j}}b_{ij,2}.
\end{align}
Finally, a decentralized second constraint is derived as
\begin{align}
	A_{i,2}u_{i} \leq b_{i,2}
	\label{eq:second-const}	
\end{align}
where
\begin{align}
	\nonumber 
	A_{i,2} = 
	\left[
	\begin{array}{c}
		A_{i1,2} \\
		\vdots \\
		A_{iN,2}
	\end{array}
	\right],
	\qquad 
	b_{i,2} = 
	\left[
	\begin{array}{c}
		\frac{U_{i}}{U_{i}+U_{1}} b_{i1,2} \\
		\vdots \\
		\frac{U_{i}}{U_{i}+U_{N}} b_{iN,2}
	\end{array}
	\right]
	\qquad
\end{align}
with $A_{ii,2} = [0, 0]$ and $b_{ii,2} = 0$.

\emph{Third constraint}: The velocity saturation is considered as the third constraint. 
Then, two ECBF candidates are designed as 
\begin{align}
	\begin{array}{l}
		h_{i,3,x} = V_{i}^{2} - v_{i,x}^{2}, \\
		h_{i,3,y} = V_{i}^{2} - v_{i,y}^{2}.
	\end{array}
	\label{eq:CBF3}
\end{align}
Using $\dot{h}_{i,3,x}  = -2 v_{i,x}(u_{i,x} - \kappa_{i}v_{i,x})$, \eqref{eq:cond-ECBF} becomes
\begin{align}
	-2 v_{i,x}(u_{i,x} - \kappa_{i}v_{i,x}) \geq -\gamma_{i,3} h_{i,3,x}
	\label{eq:const3-ECBF-cond}
\end{align}
where $\gamma_{i,3}$ are design parameters. 
Applying the inequality $2\kappa_{low}v_{i,x}^{2} \leq 2\kappa_{i}v_{i,x}^{2}$ from Assumption \ref{as:friction}, the third constraint on $u_{i}$ is obtained as 
\begin{align}
	A_{i,3}u_{i} \leq b_{i,3}
	\label{eq:third-const}
\end{align} 
where 
\begin{align}
	\nonumber
	A_{i,3} = 
	\left[
	\begin{array}{cc}
		A_{i,3,x} & 0 \\
		0 & A_{i,3,y}
	\end{array}
	\right], 
	\qquad 
	b_{i,3}  = \left[
	\begin{array}{c}
		b_{i,3,x} \\
		b_{i,3,y}
	\end{array}
	\right]
\end{align}
with $A_{i,3,m} = 2v_{i,m}$ and $b_{i,3,m} = 2\kappa_{low}v_{i,m}^{2} + \gamma_{i,3}h_{i,3,m}$; $m = x, y$.

Since the nominal controller \eqref{eq:nomin-controller} is the ideal controller to achieve the goal while not violating the game rule, the controller $u_{i}$ should be implemented as close as to the nominal controller while satisfying the three affine constraints \eqref{eq:first-const}, \eqref{eq:second-const}, and \eqref{eq:third-const}. 
Consequently, $u_{i}$ is realized by the following quadratic programming (QP)
\begin{align}
	\label{eq:QP}
	u_{i}^{*} &=  \argminC_{u_{i} \in \mathbb{R}^{2}} u_{i}^{\top}u_{i} - 2u_{i,n}^{\top}u_{i}\\
	\nonumber 
	& \mathrm{s.t.} \qquad A_{i}u_{i} \leq b_{i}
	\\ \nonumber 
	& \qquad  U_{i} \leq u_{i,x}, u_{i,y} \leq U_{i}
\end{align}
where $A_{i} = [A_{i,1}^{\top}, A_{i,2}^{\top}, A_{i,3}^{\top}]^{\top}$ and
$b_{i} = [b_{i,1}^{\top}, b_{i,2}^{\top}, b_{i,3}^{\top}]^{\top}$.

A safe set $\mathcal{S}$ for the overall multi-robot systems is now defined as
\begin{align}
	\mathcal{S} = \bigg(\prod_{i  \in \mathcal{N}}S_{i,1} \bigg) \bigcap  \bigg(\prod_{i \in \mathcal{N} }  \bigcap_{\substack{j\in \mathcal{N} \\ j \neq i}}  \mathcal{S}_{ij,2} \bigg) \bigcap \bigg(\prod_{i\in \mathcal{N}}S_{i,3} \bigg)
	\label{eq:safe-set}
\end{align}
where the product is the Cartesian product and 
\begin{align}
	\nonumber 
	S_{i,1} & = \{\{p_{i}, v_{i}\} \in \mathbb{R}^{4} | h_{i,1,x} \geq 0, h_{i,1,y} \geq 0, 
	\\ \nonumber 
	& \qquad \quad  \dot{h}_{i,1,x} + \gamma_{i,1}h_{i,1,x} \geq 0, \dot{h}_{i,1,y} + \gamma_{i,1}h_{i,1,y} \geq 0 \}, 
	\\ \nonumber 
	S_{ij,2} & = \{ \{p_{i}, v_{i}\} \in \mathbb{R}^{4} | h_{ij,2} \geq 0, \dot{h}_{ij,2} + \gamma_{i,2}h_{ij,2} \geq 0 \},
	\\ \nonumber 
	S_{i,3} & = \{ v_{i} \in \mathbb{R}^{2} |h_{i,3,x} \geq 0, h_{i,3,y} \geq 0 \}.
\end{align}

\begin{theorem}
	Given a safe set $\mathcal{S}$ for uncertain multi-robot systems \eqref{eq:robot} under Assumptions \ref{as:friction}--\ref{as:playground}, if there exist controllers $u_{i}$ satisfying \eqref{eq:QP} for $i\in\mathcal{N}$, 
	then the overall systems are safe with respect to $\mathcal{S}$.
\end{theorem}
\begin{IEEEproof}
	The proof of this theorem is similar to the proof of \emph{Theorem IV.1} in \cite{Wang2017}. 
	If all the control inputs $u_{i}$ satisfy \eqref{eq:QP}, then $S_{i,1}$, $S_{ij,2}$, and $S_{i,3}$ are forward invariant where $i,j \in \mathcal{N}$ and $j \neq i$ because of the ECBFs $h_{i,1,m}$, $h_{ij,2}$, and $h_{i,3,m}$ with $m=x,y$. 
	Thus, forward invariant $\mathcal{S}$ is verified from \eqref{eq:safe-set}, which completes the proof.
\end{IEEEproof}

\begin{remark}
	Different from the previous CBFs-based approaches for multi-robot systems \cite{Wang2017,Emam2019,Luo2020,Chen2021}, 
	the main contribution of this paper is to formulate red light, green light of multi-robot systems with friction uncertainty. 
	To handle the uncertainty while satisfying multiple safety constraints, robust inequality constraints on the control inputs have been derived using the upper and lower bounds of $\kappa_{i}$ (see \eqref{eq:const1-friction}, \eqref{eq:const2-friction}, and \eqref{eq:third-const}). 
\end{remark}

\section{SIMULATION}
\begin{figure}
 	\centering
 	\subfigure[]{\epsfig{figure=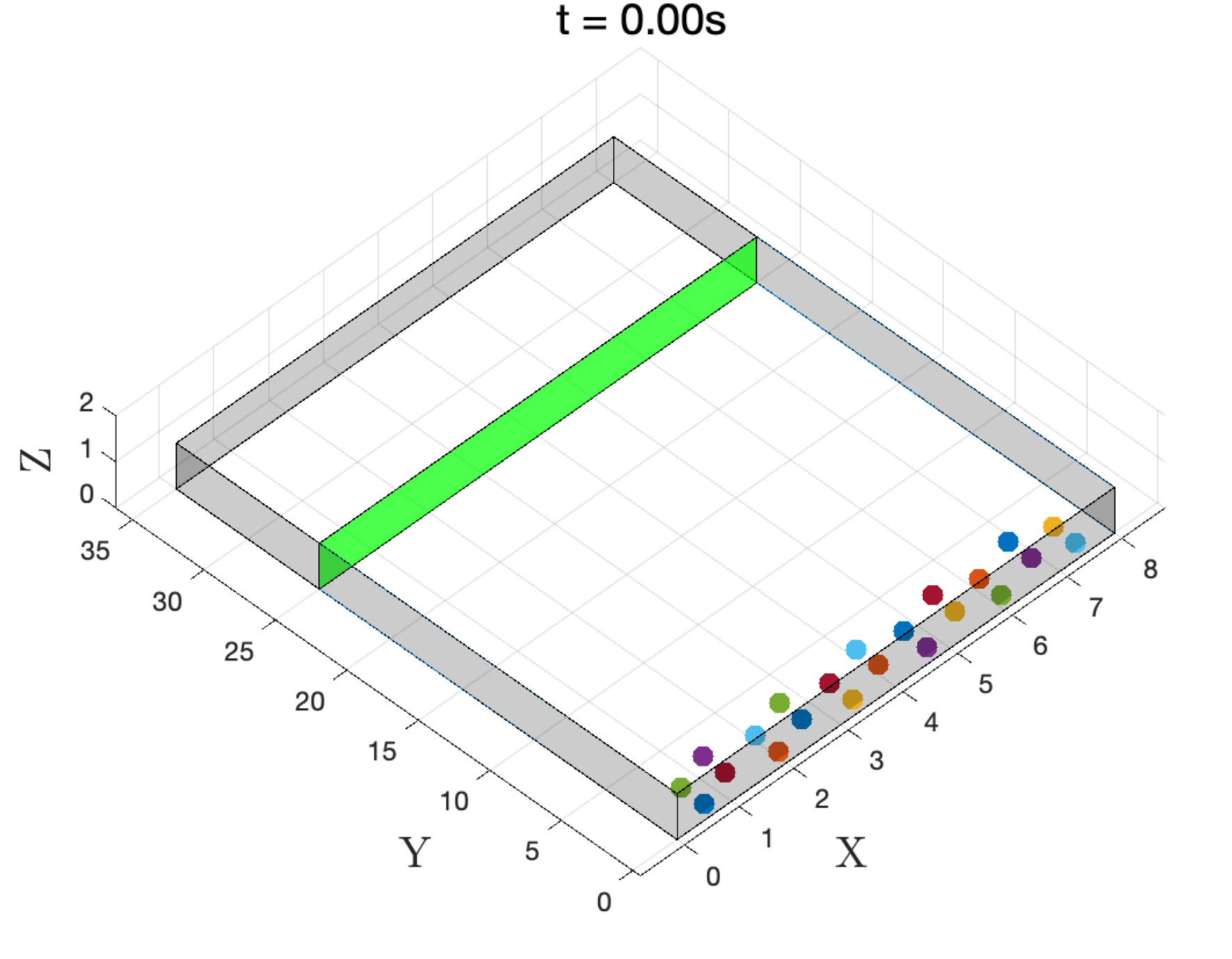, width=4.3cm,height=3.8cm}}
 	\subfigure[]{\epsfig{figure=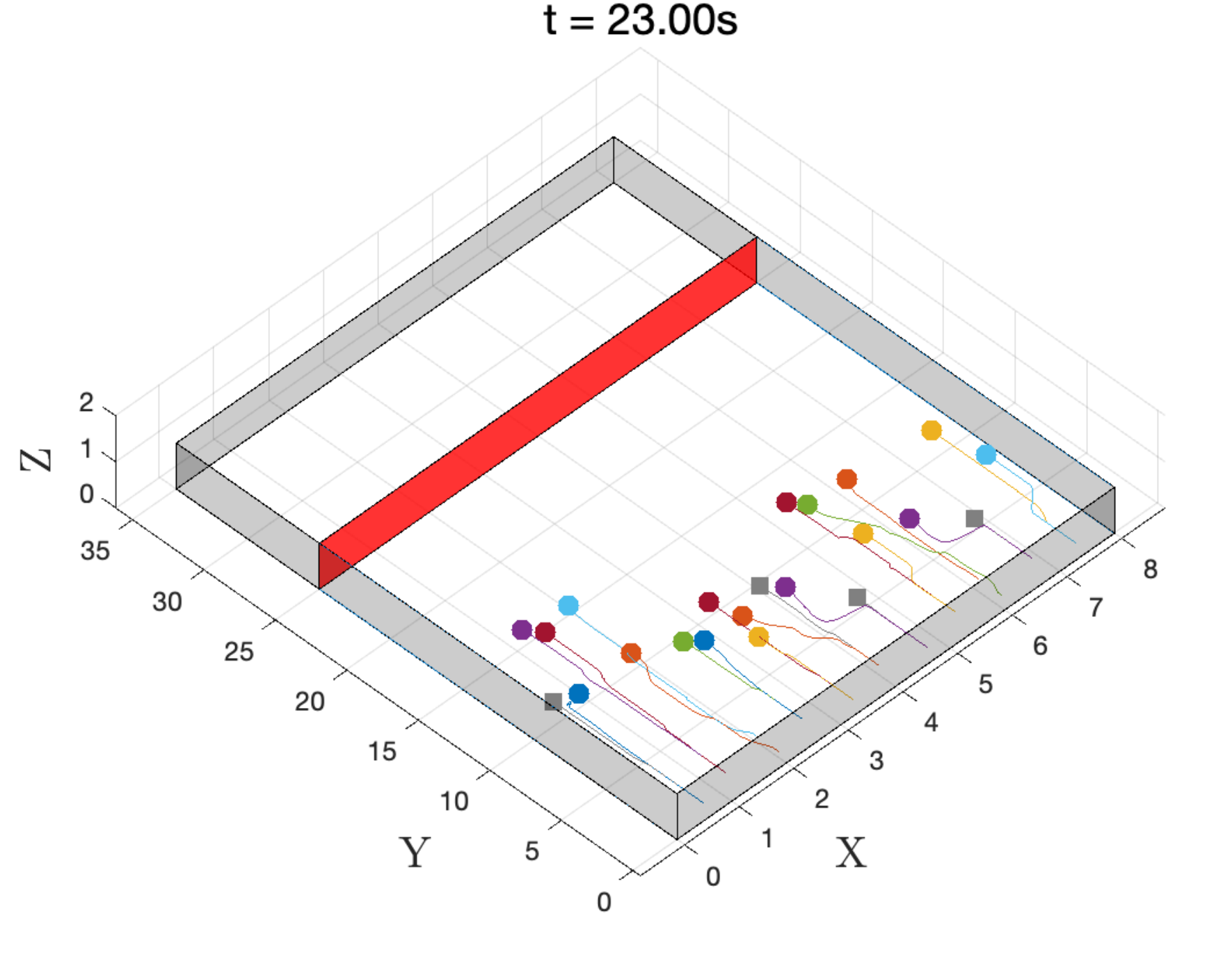, width=4.3cm,height=3.8cm}}
 	\subfigure[]{\epsfig{figure=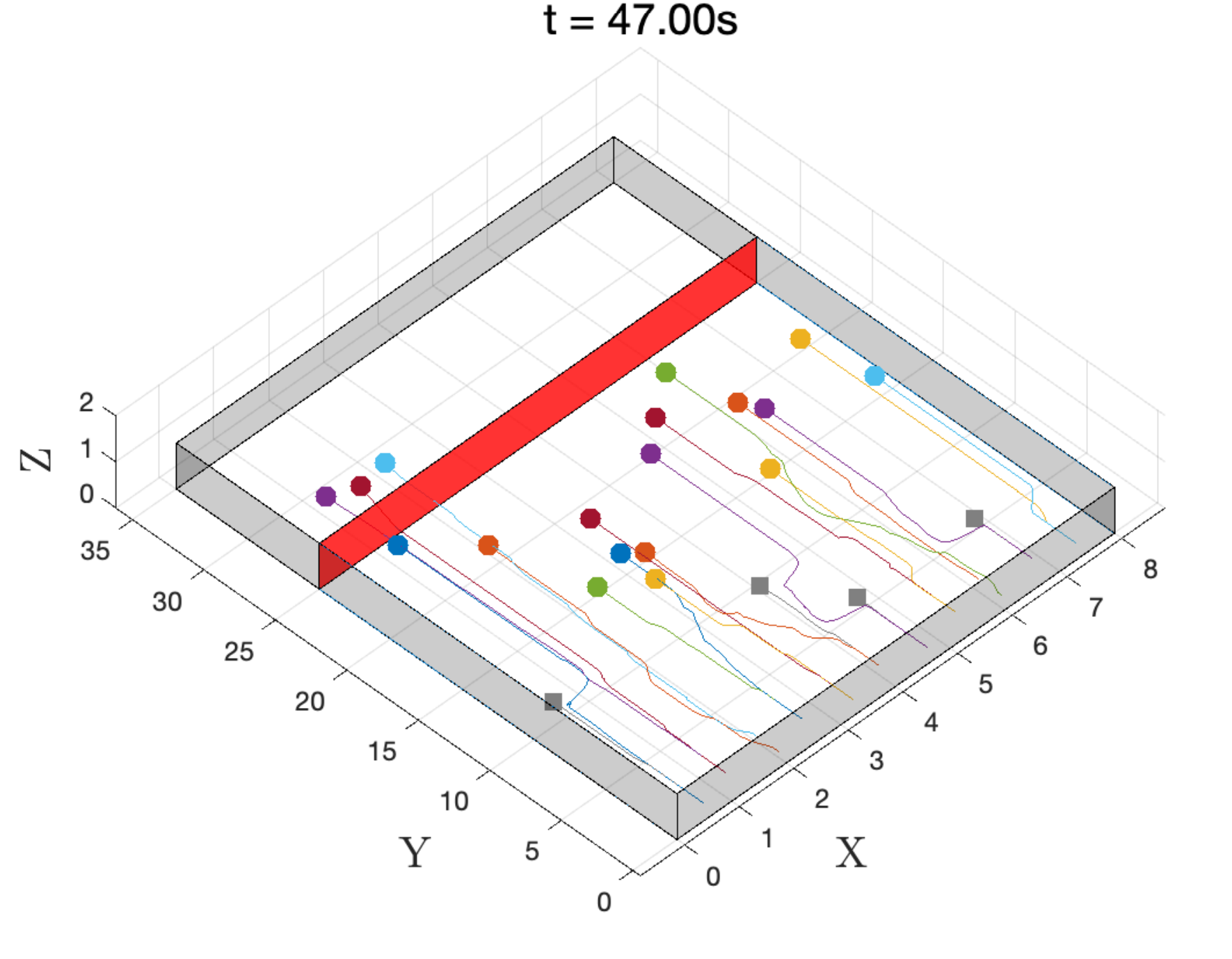, width=4.3cm,height=3.8cm}}
 	\subfigure[]{\epsfig{figure=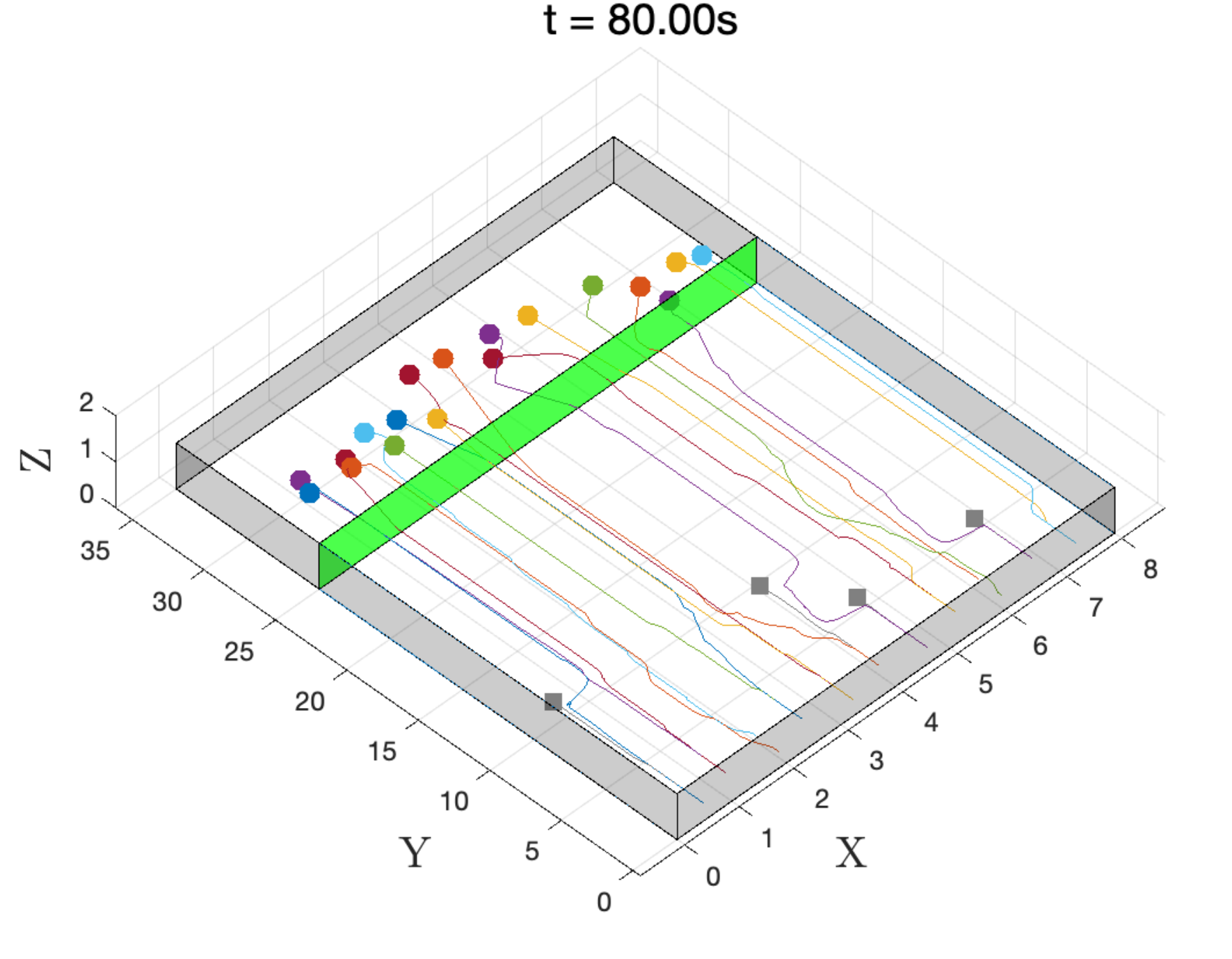, width=4.3cm,height=3.8cm}} 	
 	\caption{Four moments of multiple robots playing red light, green right when
 		(a) $t = 0$s
 		(b) $t = 23$s
 		(c) $t = 47$s
 		(c) $t= 80$s. See https://youtu.be/9AafulLU3ds for the simulation video.}
 	\label{fig:results}
 \end{figure}

 This section provides simulation results of the proposed control system for multiple robots playing red light, green light. 
 A total of $22$ robots are simulated where their initial positions are evenly spaced in the start zone of playground as illustrated in Fig. \ref{fig:results}(a) and their initial velocities are set to zeros. The game and system parameters are set to
 \begin{align}
 	\begin{array}{l}
 	t_{g_{k}} = 8(k-1), \quad t_{r_{k}} = 8k-1, \\
 	r_{0} = 0.3, \quad d_{0} = 0.4, \\
 	L_{x} = 5, \quad L_{y} = 35, \quad G_{y} = 25
 	\end{array}
 \end{align}
where $k=1,\dots,8$, $t_{g_{9}} = 64$, $t_{r_{9}} = +\infty$, and $G_{y}$ indicates the finish line. The maximum velocity and acceleration values are randomly chosen such that $1.5 \leq V_{i} \leq 2$ and $0.2 \leq U_{i} \leq 0.5$, respectively. 
The sensitivity parameters $\eta_{i}$ are also chosen as random values within $[1,1.5]$. 
For the uncertainty of the robots, unknown random friction constants $\kappa_{i}$ are considered but its upper and lower bounds $\kappa_{up}$ and $\kappa_{low}$ are known with $\kappa_{up} = 0.2368$ and $\kappa_{low} = 0.0141$. The control gain of the nominal controller \eqref{eq:nomin-controller} is set to $k_{i} = 1$ and the design parameters of the multiple ECBFs are selected as $\gamma_{i,1} = \gamma_{i,2} = \gamma_{i,3} = 5$ where $i=1,\dots,22$. 

\begin{figure}
	\centering
	\subfigure[]{\epsfig{figure=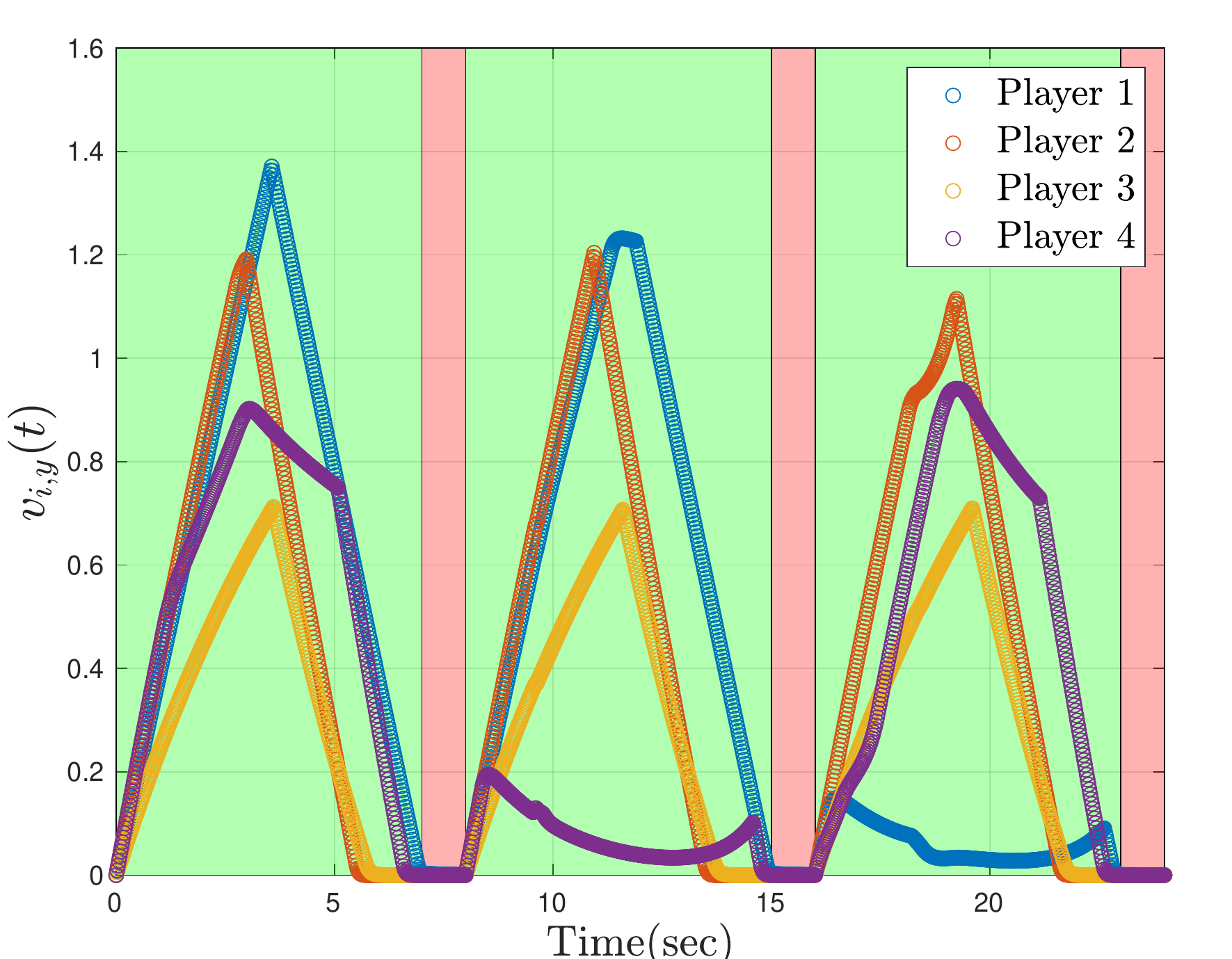, width=4.3cm,height=3.5cm}}
	\subfigure[]{\epsfig{figure=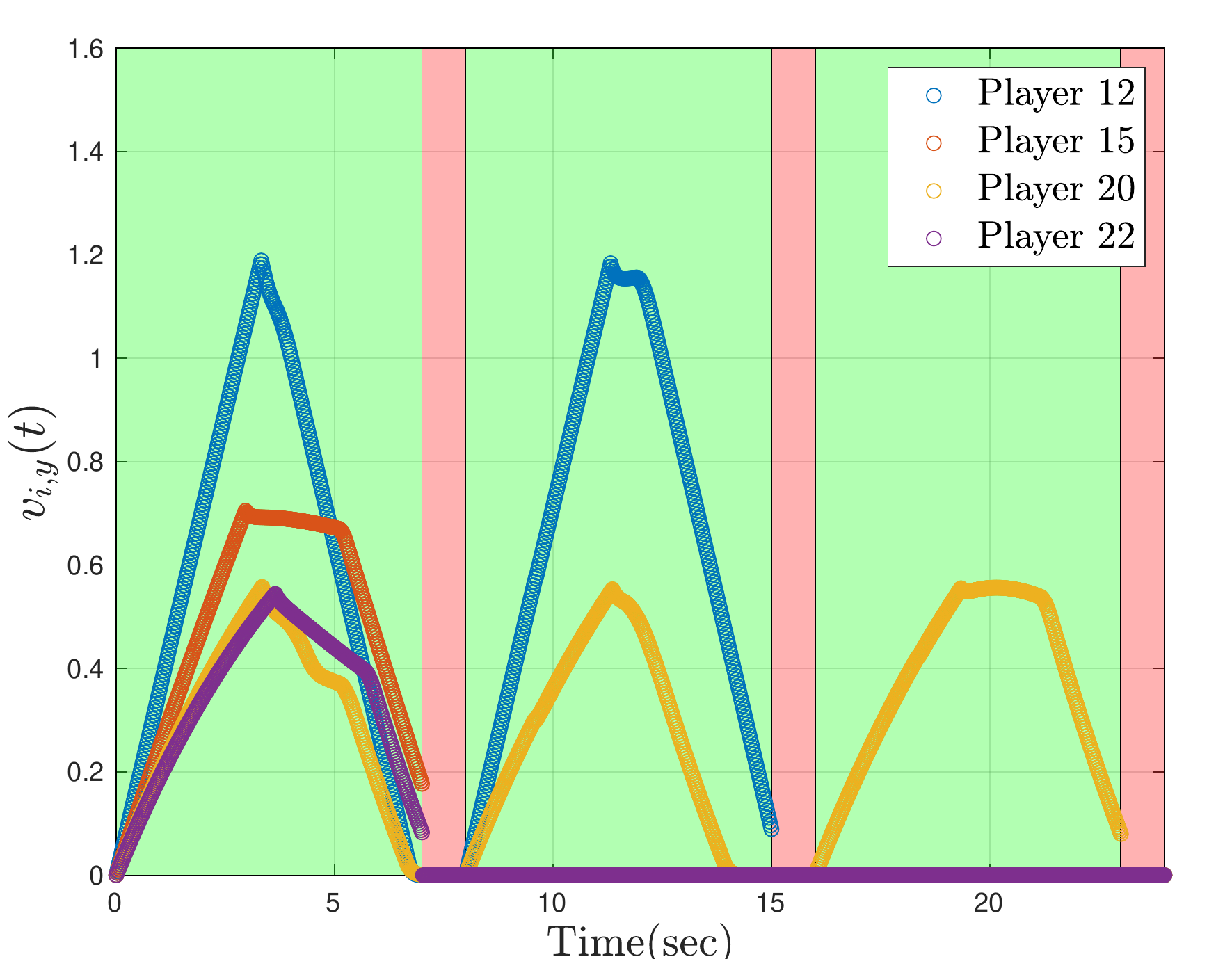, width=4.3cm,height=3.5cm}}
	\caption{Comparison of $v_{i,y}(t)$ of eight players
		(a) four live players
		(b) four dead players.}
	\label{fig:vy}
\end{figure}

\begin{figure}
 	\centering
 	\subfigure[]{\epsfig{figure=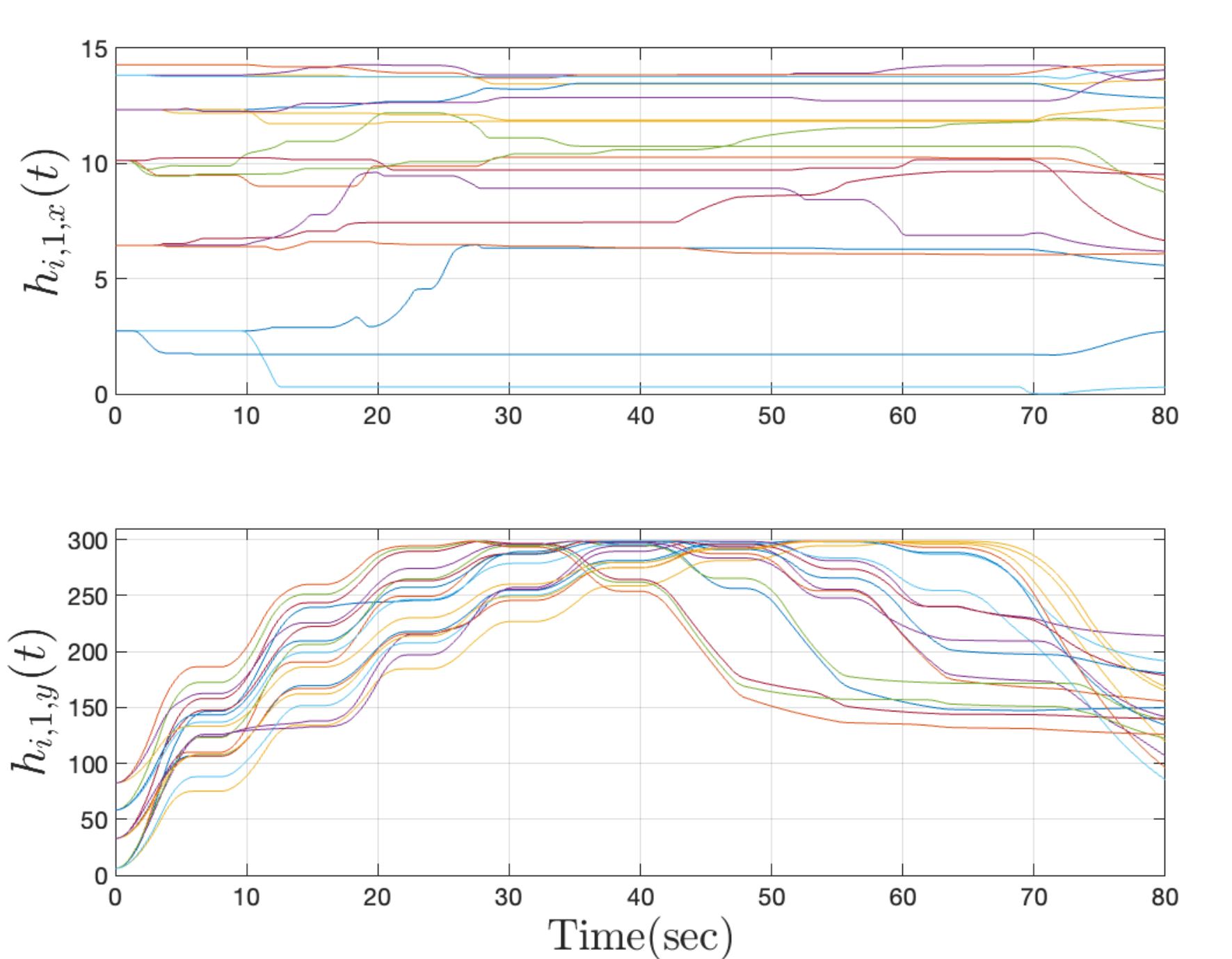, width=4.3cm,height=3.5cm}}
 	\subfigure[]{\epsfig{figure=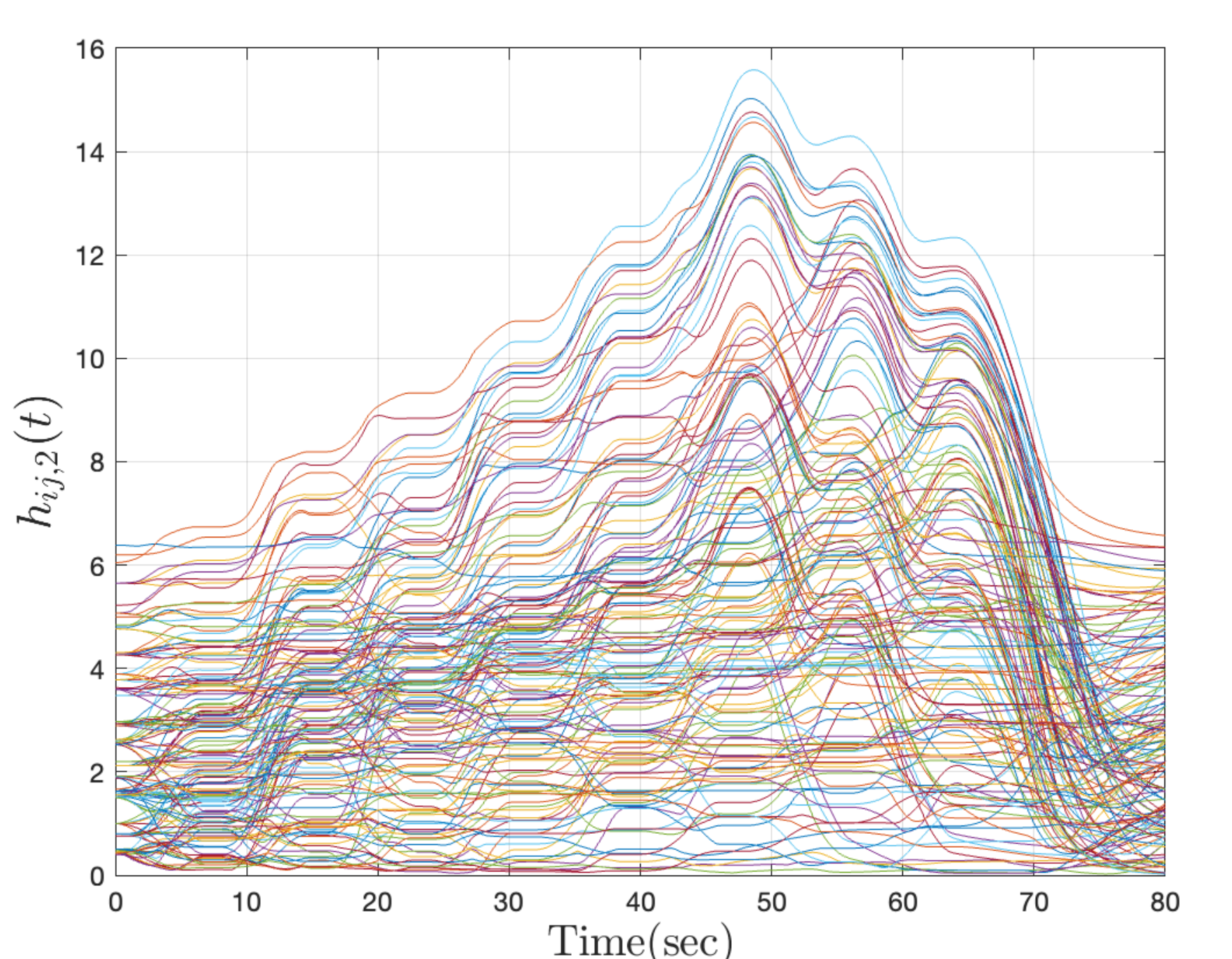, width=4.3cm,height=3.5cm}}
 	\subfigure[]{\epsfig{figure=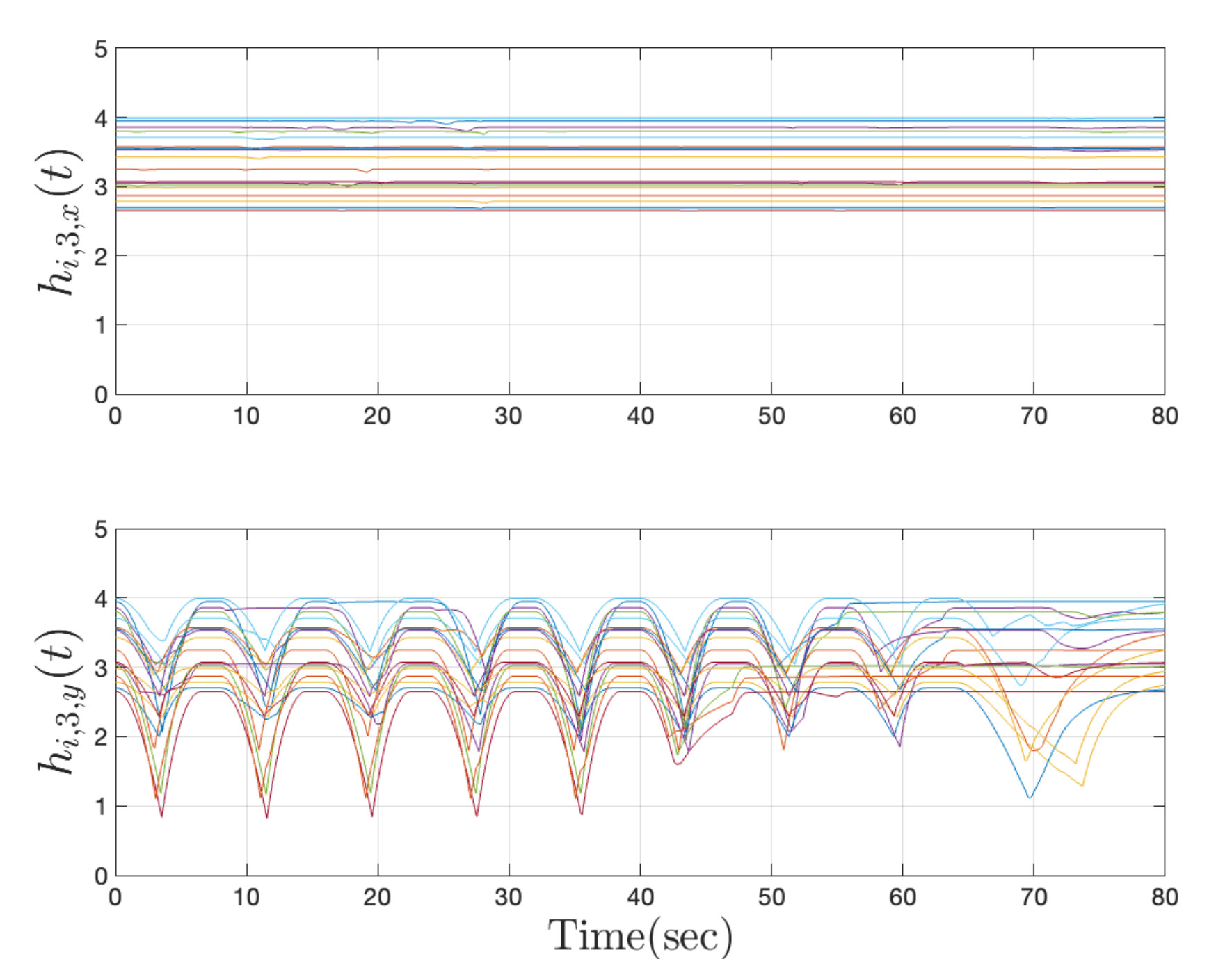, width=4.3cm,height=3.5cm}}
    \subfigure[]{\epsfig{figure=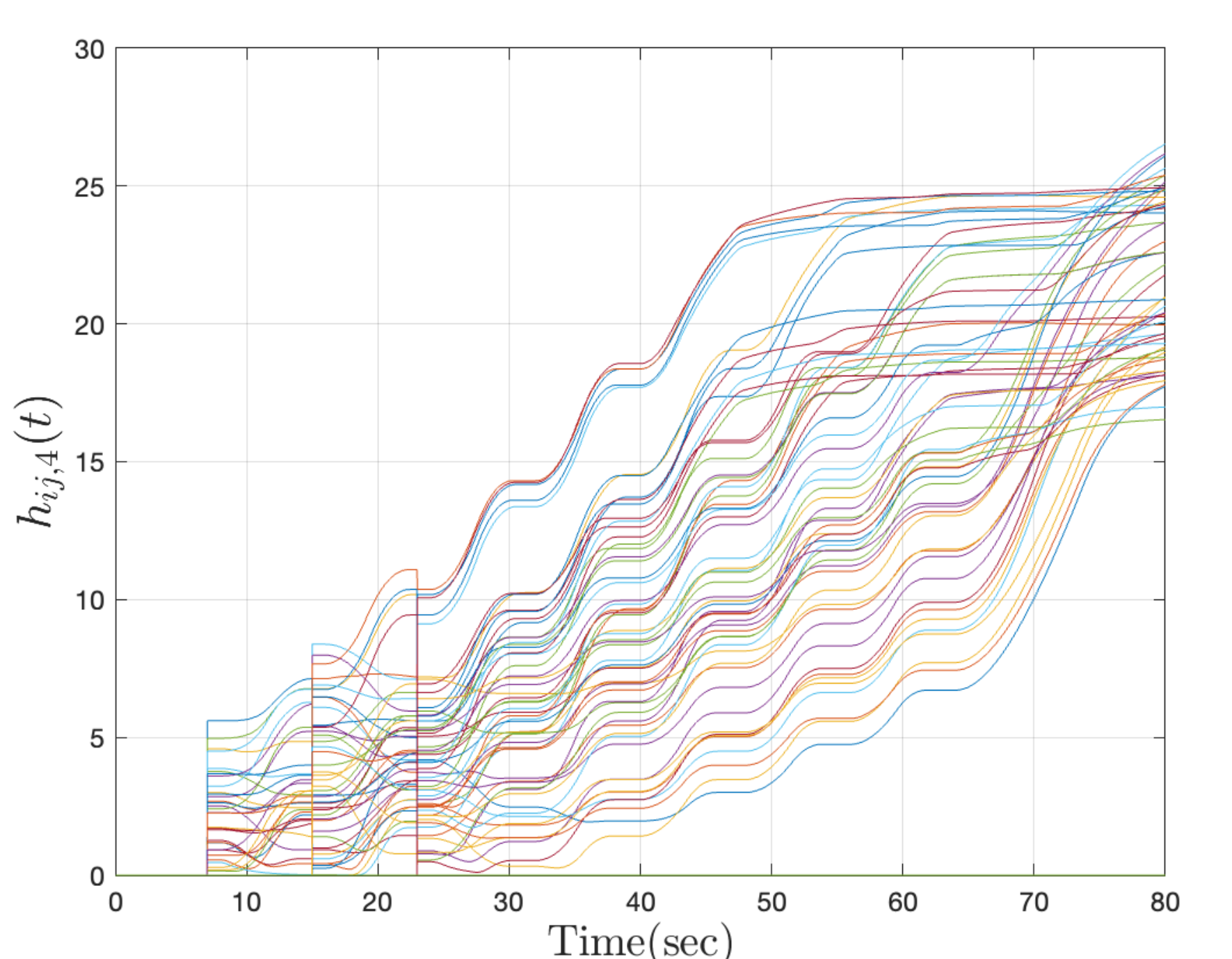, width=4.3cm,height=3.5cm}}
 	\caption{History of ECBFs 
 		(a) $h_{i,1,x}(t)$ and $h_{i,1,y}(t)$
 		(b) $h_{ij,2}(t)$
 		(c) $h_{i,3,x}(t)$ and $h_{i,3,y}(t)$
 		(d) $h_{ij,4}(t)$ for $i,j \in \mathcal{N}$ and $i \neq j$.}
 	\label{fig:h}
 \end{figure}

 In Fig. \ref{fig:results}, four moments of the multiple robots playing the game are captured 
 where the gray walls indicate the boundaries of the playground, and the colored wall represents the finish line. To show whether the robots are allowed to move or not, we change the color of the wall.
 That is, if $t \in [t_{g_{k}}, t_{r_{k}})$, the wall is green, otherwise it is red.
 In these figures, the circles mean the positions of the robots which are in play 
 and the gray squares represent the positions of the four dead robots violating the game rule. 
 Fig. \ref{fig:vy} compares the velocities on y-axis of four robots that adhere to the rule 
 with those of the dead robots. 
 As shown in Fig. \ref{fig:vy}(a), $v_{i,y}(t)$ increases first and then decreases during the green wall where $i=1,2,3$. Finally, $v_{i,y}(t)$ are zeros when the wall is red.  
 Even though $v_{i,y}(t)$, $i=12,15,20,22$, show similar histories, $v_{15,y}(t)$ and $v_{22,y}(t)$ at $t = 7$s, $v_{12,y}(t)$ at $t = 15$s, and $v_{20,y}(t)$ at $t = 23$s are not zeros. Thus, they are eliminated in the game.  
 To check the safety certificates, the ECBFs are depicted in Fig. \ref{fig:h}.
 In Fig. \ref{fig:h}, the first ECBFs $h_{i,1,x}(t)$ and $h_{i,1,y}(t)$ for playground constraints are given. From this figure, one can see that $h_{i,1,x}(t) \geq 0$ and $h_{i,1,y}(t) \geq 0$ for $t \geq 0$ which means that the playground constraints are guaranteed. 
 Similarly, the velocity constraints are also preserved by showing  
 $h_{i,3,x}(t)\geq 0$ and  $h_{i,3,y}(t)\geq 0$ in Fig. \ref{fig:h}(c). 
 The collision avoidance among robots is achieved during the play as illustrated in Fig. \ref{fig:h}(b) describing $h_{ij,2}(t) \geq 0$. 
 The last figure of Fig. \ref{fig:h} displays new ECBFs $h_{ij,4}(t)$ representing the obstacle avoidance. 
 In this simulation, we regard the dead robots (i.e., players $12$, $15$, $20$, and $22$) as obstacles after they are eliminated. Thus, the obstacle avoidance should be performed between these obstacles and the robots in play. That is why some of $h_{ij,4}(t)$ start at $t = 7$s, $t = 15$s, and $t = 23$s. The definition of $h_{ij,4}(t)$ is same as $h_{ij,2}(t)$ and thus one can easily derive inequality constraints on the control input with $u_{j}(t) = 0$ and $v_{j}(t) = 0$.
 From these figures, we can conclude that the proposed safety-critical QP controller \eqref{eq:QP} consisting of the nominal controller \eqref{eq:nomin-controller} and robust inequality constraints 
 \eqref{eq:first-const}, \eqref{eq:second-const}, and \eqref{eq:third-const} 
 can be applied to red light, green light of multi-robot systems with friction uncertainty.

 \section{CONCLUSION}
 We have addressed red light, green light game of multi-robot systems with friction uncertainty. 
 Different from the previous studies for multi-robot systems, there is a fatal condition caused by the game rule and thus we have developed a two-mode nominal controller. 
 Multiple ECBFs have been designed in order to handle several safety constraints for limited playground, collision avoidance, and saturation. While designing the nominal controller and the ECBFs, an estimated braking time and robust inequality constraints on the control input have been exploited due to the friction uncertainty. Then, a safe controller has been expressed by a QP with the nominal controller and the robust inequality constraints. 
 The proposed safety-critical controller for red light, green light game has been validated by simulation results.

\end{document}